# Is the collective IBM space exhausted only by the valence shell? *

A.D. Efimov, A.A. Pasternak, D.N. Doinikov

Cyclotron Laboratory, A.F.Ioffe Physical Technical Institute, Russia

AND

V.M. Mikhajlov

Physical Institute of St.-Petersburg State University, Russia

AND

J. Srebrny

Nuclear Physics Division, IEP, Warsaw University, Hoża 69, 00-681 Poland

Interpretation of the B(E2) values at energies higher the first backbending indicates that the maximum boson of IBM has to increase with energy and spin.

## 1. Introduction

During the last quarter of the 20th century the Interacting Boson Model (IBM) has become one of the most frequently used theoretical approaches to description of low energy collective states in atomic nuclei. It is caused in particular by the possibility of IBM to study vibrational, rotational and transitional nuclei employing the same Hamiltonian with parameters smoothly alterating along isotope or isobar chains.

As well known, in IBM1 where neutron and proton bosons are not distinguished the Hamiltonian and $T$(E2)-operator can be written in two forms. The first one [1] includes operator of scalar (s) and quadrupole (d) bosons. The second representation [2], unitary equivalent to the first one, uses square

---

roots instead of s-boson operators $d^+s \to d^+\sqrt{\Omega - \hat{n}_d}$. We write out here only $T(E2)$

$$T_\mu(E2) = e^*(d^+s + s^+d + \chi d^+d)^{(2)}_\mu =$$
$$= e'(d^+\sqrt{1 - \frac{\hat{n}_d}{\Omega}} + \sqrt{1 - \frac{\hat{n}_d}{\Omega}}\, d + \chi' d^+d)^{(2)}_\mu,$$

where $\Omega$ is the maximum boson number. The finiteness of the boson number is the postulate of IBM which is grounded on the supposition that *each d-boson is a boson image of a quadrupole nucleon pair* occupying levels of the valence shell in a given nucleus. Therefore the d-boson number can not exceed the half of the number of valence nucleons or holes ($\Omega_{\text{st}}$).

It is worthwhile to stress that in IBM the standard choice of $\Omega$, $\Omega_{\text{st}}$, is not a result of some general principles. However, if in forming the collective space take part states of adjacent shells, then $\Omega$ can be taken higher and its value may be varied depending on energy and spin of states under consideration. Thus $\Omega$ can be regarded as one more parameter of the model and its empirical determination is of interest as it can display *what part of the fermion space is covered by the collective space.* This question was investigated for deformed nuclei in Ref. [3]. Here we discuss the choice of $\Omega$ mainly in transitional nuclei.

## 2. $\Omega$ dependence of energies and B(E2) values

The lowest eigen values of IBM Hamiltonian are rather weakly sensitive to the choice of $\Omega$. In fact, in the $SU_5$ limit the ideal oscillator spectrum ($E = \varepsilon n_d$) is independent of $\Omega$ at all. Analogously in the $SU_3$ limit the low energy spectrum is purely rotational $E = I(I+1)/2J$, $I$ is the angular momentum, and depends on the only parameter, the moment of inertia. The grounds of the $\beta$- and $\gamma$-bands depends on $\Omega$ explicitly. Nevertheless a variation of $\Omega$ can be compensated by a corresponding variation of the Hamiltonian parameters and effective charge. In more general case the $\Omega$ independence of wave functions and eigen values is conserved if the expectation value of $n_d$ does not significantly exceed the value $I/2$. The termination of the collective energy spectrum at $I = 2\Omega$, that is a strong prediction of IBM, cannot be really verified as this state lies so high in energy that it cannot be separated from other high spin levels created by two- and four-quasiparticles.

More evident information concerning $\Omega$ is given by the B(E2) values for transitions along the yrast band. IBM predict that B(E2) vs. $I$ maximum at $I \approx \Omega$ and then begins to diminish. This prediction is realized in all limits of IBM that drastically distingushes IBM from geometric models.

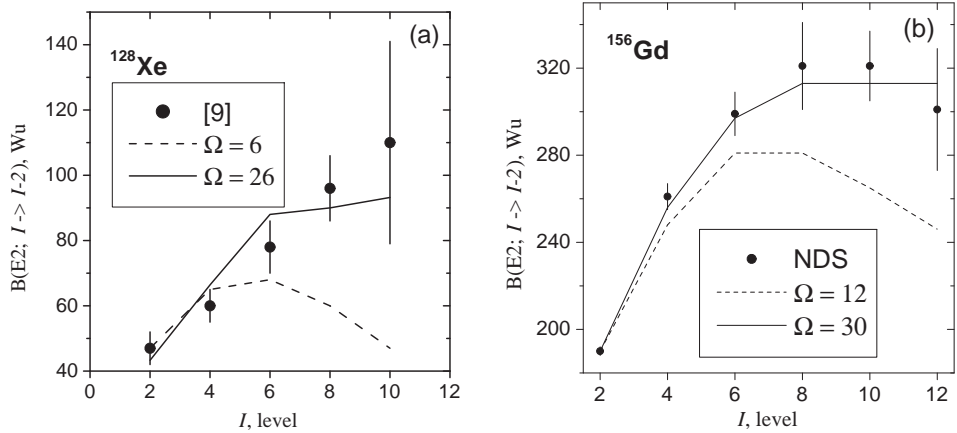

Fig. 1. (a) The comparison of experimental B(E2) values in $^{128}$Xe [9] with IBM calculations: solid and dash lines respectively correspond to $\Omega$=26 and 6.
(b) The experimental B(E2) values in $^{156}$Gd in comparison with IBM calculations with $\Omega$=12 (dash line) and in the SU$_3$ limit with $\Omega$=30 (solid line).

Thus, one of the ways for empirical determination of $\Omega$ could consist in studying the yrast B(E2)'s. Then that value of $I$ at which B(E2) is maximum could be taken as an estimation for $\Omega$. There exist several nuclei, e.g. $^{74}$Se [4], where such maximum is rather clearly pronounced at $I \approx \Omega_{\text{st}}$.

Nevertheless diminishing B(E2) after some value of $I$ does not determine $\Omega$ unambiguously since the crossing of collective and high spin two-quasiparticle bands can also lead to attenuation of B(E2). First such behaviour of B(E2) was considered in Ref. [5] where the band crossing was considered by using the known direct mechanism of the quasiparticle-phonon interaction [6]. However, there exist exchange mechanisms of the quasiparticle-phonon interaction [7] which increase the mixing and simultaneously B(E2) values in the crossing region. Nevertheless all considered ways of mixing result in decreasing B(E2), and they cannot explain increasing B(E2) after $I > \Omega_{\text{st}}$. Therefore examples of the growth of B(E2) after $I = \Omega_{\text{st}}$ for transitional and deformed nuclei $^{100}$Mo [8], $^{128}$Xe [9] and $^{156}$Gd testify to that the rule $\Omega = \Omega_{\text{st}}$ has to be revised. The B(E2) values for $^{128}$Xe and $^{156}$Gd are compared with IBM calculations in Figs. 1 for $\Omega_{\text{st}}$ and $\Omega > \Omega_{\text{st}}$.

## 3. Bandcrossing and increasing B(E2) values in $^{120}$Xe and $^{118}$Te

Spectra of many Te, Xe, Ba, Ce nuclei possess some common features that explicitly indicate the interaction between the collective quadrupole mode and the high-spin two-quasiparticle excitations involving h$_{11/2}$ quasi-particles. This interaction leads to a fragmentation of the collectivity over several high-spin states. In Ref. [7] level structure of nuclei in this mass region were analyzed with a version of IBFM in which a possible shape instability was taken into account by means of variations of collective operator parameters. With increasing energy and spin the structure of the collective quadrupole excitations can change. The excitation of high spin two-quasiparticle pairs intensifies this change, narrowing the collective space and bringing in additional variations in the mean field and pairing. Phenomenologically we take these alterations into account by using different sets of IBM parameters in boson images of operators. However we perform detailed microscopic considerations of the interaction between collective quadrupole phonons. The phonon structure and their energies are calculated in the framework of RPA.

The comparison of the theoretical and experimental energy spectra for $^{120}$Xe was presented in Ref [10]. For the yrast states, $I \leq 10^+$, the collective components prevail ($\geq 60\%$). On the contrary, the states with I$\geq$14 are built on $10^+$ $\nu(h_{11/2})^2$ excitations. The $12^+$ state is transitional: s-, d- collective and two-quasiparticle components are uniformly mixed. To explain the experimental observations B(E2) presented in Fig. 2a we suppose that the maximum boson number of IBM can depend on the energy and structure of the excited states. The best description of B(E2) is attained by using in the purely collective components of the wave function the value of $\Omega_C = \Omega_{st} - 2$ and in the components including two-quasiparticle pairs $\Omega_{qp} + 1 = \Omega_{st} + 10$. This approximation especially appropriate for high-spin states indicates that the maximum boson number of IBM may increase with energy and spin.

The B(E2)-values of $^{118}$Te measured in the work [14] give us the unique possibility to analyze E2 transition probabilities in the collective band above the band crossing. Usually in this spin region we deal only with transitions between states involving two-quasiparticle pairs. Our calculations in the framework of IBFM reproduce quite reasonably the energy spectrum of $^{118}$Te up to 7 MeV, however the employment of the standard IBM E2 operator leads to too small B(E2) values in comparison with experimental ones. Since the IBM attenuation begins to work when $n_d$ exceeds $\approx \Omega/2$ for explanation of the experimental B(E2) values shown in Fig. 2b, we are forced to choose $\Omega$ larger than $\Omega_{st}$: $\Omega_C = \Omega_{st} + 10$ and $\Omega_{qp} + 1 = \Omega_{st} + 10$.

## 4. Conclusion

In the present work we interpret the experimental data on E2-transitions supposing that the boson number $\Omega$ can increase with excitation energy and spin that gives evidence of extension of the collective space outside the valence shell. The first reason for such extension is complicated structure of the collective quadrupole phonon, a fermion counterpart of the d-boson of IBM, comprising excitations inside a valence shell and high energy particle-hole pairs which may essentially contribute to the many boson wave function. The second reason consists in the mutual influence of the quadrupole excitations and the mean field that gives rise to deformations reducing shell energy gaps and extending thereby the collective space.

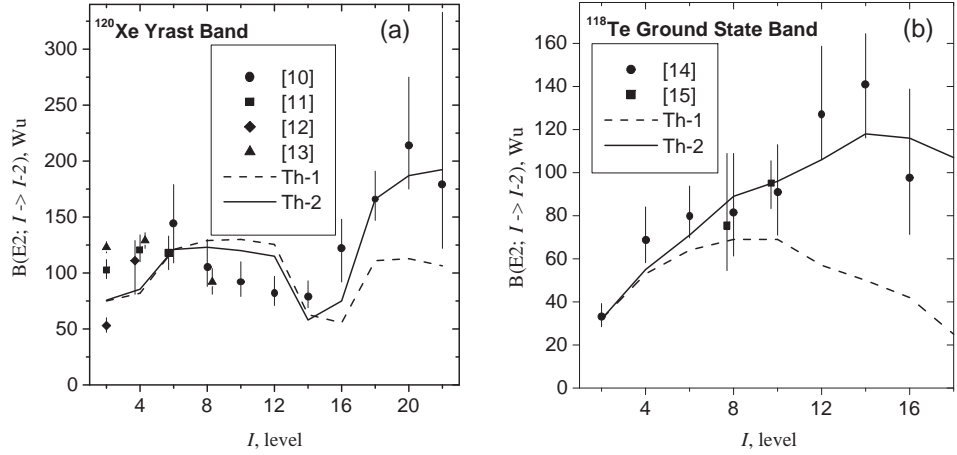

Fig. 2. (a) The experimental B(E2) values for transitions along the yrast band in $^{120}$Xe. Calculations are performed with $\Omega_c = \Omega_{qp} + 1 = \Omega_{st}$ (dash line — Th-1) and $\Omega_c = \Omega_{st} - 2$, $\Omega_{qp} + 1 = \Omega_{st} + 10$ (solid line — Th-2); $\Omega_c$, $\Omega_{qp}$ are the maximum boson number in purely collective states and in those with a quasiparticle pair, $\Omega_{st}$ is the standard IBM value determined by the half of the valence nucleon or hole number.
(b) The comparison of experimental B(E2) values inside the ground state band in $^{118}$Te with calculations at different values of $\Omega$: $\Omega_c = \Omega_{qp} + 1 = \Omega_{st}$ (dash line — Th-1) and $\Omega_c = \Omega_{qp} + 1 = \Omega_{st} + 10$ (solid line — Th-2).